# Bend it like dark matter!

**J Woithe[1], M Kersting[2]**

[1] CERN, Geneva, Switzerland

[2] Department of Teacher Education and School Research, University of Oslo, Oslo, Norway

E-mail: julia.woithe@cern.ch

October 2020

**Abstract.** Dark matter is one of the most intriguing scientific mysteries of our time and offers exciting instructional opportunities for physics education in high schools. The topic is likely to engage and motivate students in the classroom and allows addressing open questions of the Standard Model of particle physics. Although the empirical evidence of dark matter links nicely to many standard topics of physics curricula, teachers may find it challenging to introduce the topic in their classrooms. In this article, we present a fun new approach to teach about dark matter using jelly lenses as an instructional analogy of gravitational lenses. We provide a brief overview of the history of dark matter to contextualise our presentation and discuss the instructional potential as well as limitations of the jelly lens analogy.

## 1. Introduction: Why should we teach about dark matter?

Dark matter is one of the most intriguing scientific mysteries of our time. While this invisible type of matter seems to be abundant in our cosmos, physicists grope in the dark about its nature and origin. Attempting to verify dark matter directly, scientists investigate possibilities of physics beyond the Standard Model. Recently, an unexpected signal from the dark matter detector XENON1T set off a wave of excitement among dark matter hunters (1). Likewise exciting are the instructional opportunities of dark matter in the classroom. Not only does the topic engage and motivate high school students while introducing them to fundamental concepts of particle physics, astronomy, and cosmology. The search for dark matter is also an excellent example of science in the making which illustrates scientific practices and aspects of the nature of scientific knowledge.

In this paper, we motivate our interest in teaching dark matter by providing a brief overview of its instructional potential in the classroom. We then turn to the empirical evidence that has led physicists to believe that dark matter exists before presenting different ways of teaching about dark matter based on this evidence. Employing jelly lenses as an analogy of gravitational lenses, we describe an easy and fun experimental setup for classroom instruction. We then discuss various use cases as well as limitations of the jelly lens analogy to support teachers in their endeavour of introducing dark matter in the classroom.

*1.1. The instructional potential of dark matter*

The topic of dark matter has great instructional potential that relates both to the physics content and students' motivation in physics:

*Motivation and student engagement* By its very nature, dark matter can much motivate and engage students. Dark matter is a topic of modern physics, it features prominently in our understanding of the universe, and it presents scientists with unsolved problems. Research has shown that all these factors can have a positive impact on students' motivation and their attitudes towards science. For example, the large-scale survey ROSE (Relevance Of Science Education) found topics of physics in space and astronomy to be popular among 15-year-old students (2). The survey also found that students were interested in seemingly mysterious scientific phenomena or phenomena that scientists cannot yet explain. Other studies have corroborated these findings, which seem to be robust irrespective of age, gender or nationality (3,4). Inviting students to explore complex phenomena and unsolved problems in modern physics can provide productive learning opportunities (5–7).

*Curriculum links* Despite its mysterious nature, dark matter is a very accessible topic at the high school level. Teachers can link dark matter to many concepts that are already part of standard physics curricula such as gravity, circular and orbital motion, and optics. There are obvious links to topics of astronomy, particle physics and general relativity that have started to become an increasing focus of curricula reforms in recent years (8,9). Hence, dark matter provides teachers with flexibility in choosing how to integrate the topic into existing practices.

*Nature of science* Dark matter is an excellent example to illustrate "science in the making" because there are many competing theories about the origins and the composition of dark matter. Teaching about dark matter can foster awareness of the tentative nature of scientific knowledge and illustrate aspects of scientific practices. There is a growing awareness in the physics education community that such familiarity with scientific methods enables students to contextualise their physics learning in a broader societal context (10,11). Thus, dark matter provides opportunities to embrace broader methodological, epistemic, and social perspectives of physics.

In summary, dark matter can serve as a hook to get students interested and engaged, it links to standard topics in physics curricula, and it allows teachers to incorporate perspectives of the nature of science.

## 2. Background: What is dark matter and how can we bring this topic into classrooms?

We frame our approach to teaching about dark matter in line with the Model of Educational Reconstruction (12). The model can guide the development of instructional approaches about dark matter by integrating three components of science education research and practice. The first step involves identifying key concepts of the science content. In a second step, students' perspectives are taken into account, often in the form of common (mis)conceptions in this learning domain. Finally, an instructional approach is designed that combines both science content and student conceptions. This final step includes careful considerations of the strengths and limitations of the chosen instructional model. In this section, we provide a brief overview of all three components of the Model of Educational Reconstruction with a view towards teaching dark matter. We hope that this overview is interesting and useful to teachers of physics at the high school level. Our particular focus lies on spacetime curvature and gravitational lensing. These empirical clues rely on the general theory of relativity and are, therefore, more abstract than the classical empirical evidence of dark matter.

*2.1. Science content structure: what is dark matter and how do we know it's there?*

How do we know something is there if we do not see it? For centuries, physicists and astronomers have relied on visual observations to map the cosmos. However, visual observations depend on electromagnetic interactions. If something does not interact electromagnetically, it remains invisible to the naked eye and our telescopes. Today, most scientists believe that about 85% of all matter in the universe is indeed invisible. No one knows the true nature of this dark matter. Still, there is overwhelming indirect evidence that large amounts of dark matter are present. While dark matter cannot be observed directly, it has visible effects due to its gravitational interaction with ordinary visible matter. Over the last century, physicists and astronomers have gathered different clues and observations that have provided convincing evidence that there is more in our universe than meets the eye. Box 1 gives a brief overview of the classical and modern empirical evidence of dark matter that makes use of classical mechanics and general relativity, respectively.

> **Box 1: Classical and modern empirical evidence of dark matter**
>
> - **Something is missing**: In the 1930s, Swiss astrophysicist Fritz Zwicky studied the Coma Cluster and applied the virial theorem of classical mechanics to this cluster of galaxies. Estimating the mass of the cluster based on the motion of its galaxies and comparing this to an estimate of mass based on the brightness of the galaxies, Zwicky inferred the existence of unseen dark matter that held the cluster together.
>
> - **Something rotates weirdly**: In the 1970s, American astronomer Vera Rubin and her colleagues confirmed Zwicky's missing-mass problem by providing independent evidence of dark matter in the form of galactic rotation curves. The arms of spiral galaxies rotate around their galactic centre, and Kepler's Second Law predicts a decrease in the rotation velocities towards the edges of the galaxy. However, Rubin discovered a discrepancy between the predicted and the observed angular motion of these spiral galaxies. She concluded that there must be a lot of dark matter in the outskirts of the galaxies.
>
> - **Gravitational lensing**: In the 1990s, scientists gathered further evidence for the existence of dark matter that relied on the gravitational bending of light. According to general relativity, massive objects curve spacetime. The curved geometry of spacetime acts like a lens by deviating the path of light that distant sources emit. Observers on Earth see distorted images of these light sources, and by measuring the amount of distortion, they can estimate the masses of the gravitational lenses. The great advantage of this method is that it provides a way of estimating the masses of objects without making assumptions about the nature of the mass (i.e. ordinary or dark matter).
>
> - **Bullet cluster**: In the early 2000s, observations of the Bullet cluster provided our current-best evidence for dark matter yet. The Bullet cluster consists of two merging galaxy clusters in which hot gas makes up most of the ordinary matter. If dark matter did not exist, one would expect gravitational lensing phenomena to follow the distribution of the gas. However, gravitational lensing studies of the merger revealed the centre of mass to be spatially displaced from the hot gas. This observation suggests that large amounts of dark matter exist in both galaxy clusters which did not interact with the gas during the collision.
>
> - **On-going search**: Currently, most scientists believe that undiscovered particles are the source of dark matter, and many experiments such as XENON1T (1) aim to detect these particles. Among the most promising candidates for dark matter particle are the so-called WIMPs, weakly interacting massive particles.

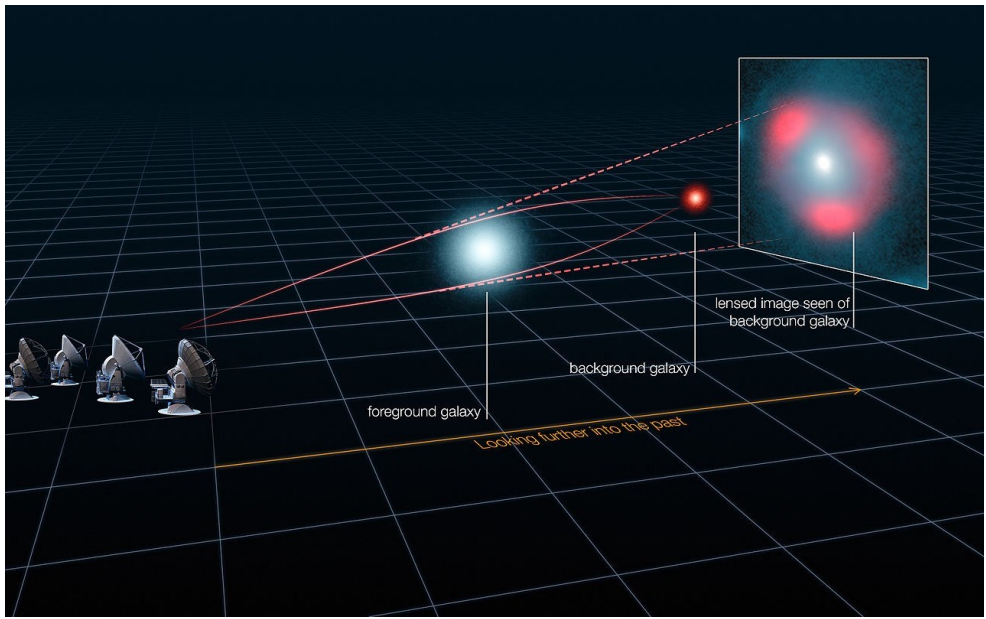

**Figure 1.** The gravitational lensing effect stems from geometric distortions in spacetime: A foreground galaxy curves spacetime and, thereby, acts as a "lens" by deflecting the path of light of a distant background galaxy. On Earth, telescopes capture the distorted image of the background galaxy from which scientists can infer the mass of the foreground galaxy. Credit: ALMA (ESO/NRAO/NAOJ), L. Cal¸cada (ESO), Y. Hezaveh et al., CC BY 4.0 https://creativecommons.org/licenses/by/4.0, via Wikimedia Commons

*Gravitational lensing* In this paper, we focus on gravitational lensing as an instructive empirical clue of dark matter. As opposed to the classical empirical evidence of dark matter, gravitational lensing cannot be explained by classical physics alone. To make sense of this phenomenon, we need Einstein's general theory of relativity according to which massive objects such as stars and galaxies curve spacetime. Light propagates through spacetime along straight lines. Consequently, if the geometry of spacetime is curved, this affects the trajectories of light as well. Light continues to follow the straightest possible paths through curved spacetime which we interpret as the gravitational deflection, or gravitational lensing, of light (Figure 1).

The terminology of gravitational lenses stems from an analogy with ray optics. Just like optical lenses deflect light from its straight trajectory because of their refractive power, so does a "gravitational lens" exert such a massive gravitational effect that the propagation of light is visibly influenced. Telescopes can capture such bending or distortions of light when massive objects act as gravitational lenses (Figure 2).

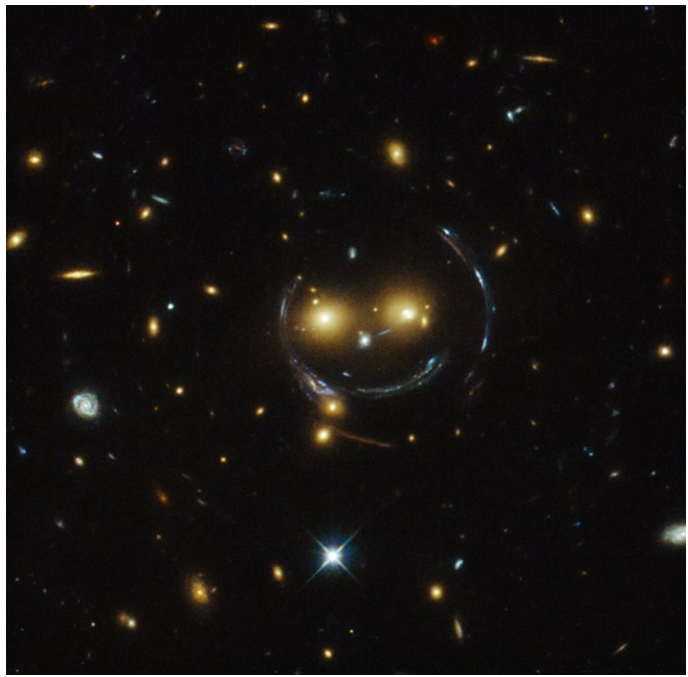

**Figure 2**. The NASA/ESA Hubble Space Telescope took this image of two faint galaxies that seem to be smiling: the two eyes are the galaxies SDSSCGB 8842.3 and SDSSCGB 8842.4 and the smile lines are arcs caused by strong gravitational lensing. Credit: ESA/Hubble, CC BY 4.0 https://creativecommons.org/licenses/by/4.0, via Wikimedia Commons

How do we know a gravitational lensing effect was caused by dark or ordinary matter? We don't! Indeed, for spacetime to curve, it does not matter whether the massive object is ordinary matter, dark matter or even a black hole. Only the amount of mass and its distribution determines spacetime curvature and thus, the gravitational lensing effect. And this independence of the composition of matter has made gravitational lenses crucial in the search for dark matter because they provide an independent measurement of the mass of cosmic objects. For example, in addition to the classical evidence of dark matter, gravitational lenses offer a way of estimating the mass-to-light ratios of cosmic objects that uses assumptions of general relativity. The measurement of these ratios based on data from gravitational lenses has confirmed the classical missing-mass problem. Analyses of patterns of multiple lens images have also provided further clues about the actual distribution of dark matter in galaxies and galaxy clusters.

*2.2. Student conceptions: how do students think about dark matter, spacetime, and optical lensing?*

To find instructional approaches for teaching dark matter that are tailored to the needs of high school students, we need to consider students' conceptional difficulties in this learning domain. In the specific case of gravitational lensing, we have to take three domains into account:

*Dark matter* The amount of previous research on high school student conceptions of dark matter is minimal. However, Coble et al. (13) reported that even undergraduate students were mostly unaware of the existence and origins of dark matter when entering a general education astronomy course. Moreover, even after instruction, students had difficulties in distinguishing between dark matter and dark energy. Finally, the authors suggest that students' maths skills, and in particular, the ability to read graphs, interfered with understanding the rotation curves of spiral galaxies as evidence for dark matter. This is an important finding since insufficient familiarity with reading graphs is likely to impede high school students' understanding of the subject as well. Based on numerous informal discussions with students and teachers that we have had in our role as physics educators, we wish to add another observation: Students often struggle to distinguish dark matter from dark energy but also from other "mysterious" concepts such as antimatter and black holes (14).

*Curved spacetime* In the context of gravitational lensing, students need to build understanding both of curved spacetime and light propagation in spacetime. Similar to the concept of dark matter, empirical evidence about high school students' understanding of spacetime is limited. First evidence suggests that high school students were intrigued by the idea of spacetime but struggled to visualise it (8,9). The four-dimensional nature of spacetime seemed to pose conceptual difficulties when students tried to conceptualise curvature or the geometry in four dimensions. Students were particularly bewildered about the time dimension and how time could be "curved" or "warped" (15). We suggest using students' curiosity about spacetime as a hook to address our inability to visualise more than three dimensions. Encouragingly, high school students seemed to understand the concept of straight paths in curved spacetime pretty well. According to general relativity, light travels along straight trajectories (so-called geodesic curves) through spacetime. High school students appear to be able to characterise geodesic curves as the curved-spacetime generalisation of a straight line in various different but equivalent ways (16). This finding is encouraging since it suggests that students can deal with the conceptual foundation of gravitational lensing.

*Ray optics* In a study with high school and teacher training students, Galili and Hazan (17) described several patterns of reasoning in optics that deviated from a scientifically correct interpretation. For example, the authors referred to the "flashlight scheme" if learners thought individual light rays were emitted from single points on the light source. Moreover, students tended to consider and draw only the relevant light rays emitted towards the target object (e.g., a lens) while ignoring light emitted in other directions. The authors referred to the "image holistic scheme" if students interpreted an image as a material replication of an object. For example, some students argued that images could only be observed if they enter the eye of the observer.

Moreover, the "spontaneous vision scheme" summarised student conceptions of seeing as a process that happens naturally in the presence of the eye. Indeed, some students thought that light was only needed to illuminate an object, or the eye itself. The students were often unaware of the role of light in the seeing process and the role of the eye as a photon detector.

Consequently, some students thought that light rays could be seen from the side, a misconception that is often supported by illustrations of lensing effects. Finally, for triangular prisms, students thought that the prism created two images by splitting up the light rays originating from the light source. This misconception is particularly relevant for us since our jelly lens analogy builds on a triangular prism construction.

*2.3. Design of instructional resources: how do we teach about dark matter?*

Dark matter can seem like a very abstract field of research, and its mathematical foundation is usually far too complicated for high school students. Nevertheless, inspired by the instructional potential of dark matter, educators from different research institutions provide inquiry-based learning activities for high school students. These learning activities focus on different aspects of dark matter research (see Box 2). Moreover, since 31 October 2017, research institutions from around the world have celebrated "dark matter day" to highlight current dark matter research [18]. We take dark matter day as an opportunity to introduce a new and fun approach to teaching about dark matter that makes use of jelly lenses.

> **Box 2: Overview of existing instructional activities on dark matter**
>
> - **Something is missing:** NASA JBL propose an activity where students make observations of two plastic bottles, both filled with small objects such as washers. One bottle is also filled with water. Although the water is invisible, students can observe the different masses of the two bottles (18).
> - **Something rotates weirdly:** A lesson plan offered by the Sanford Underground Research Facility introduces a weirdly spinning paper plate. Here, students infer the existence of extra hidden mass in the form of washers (19). The Perimeter Institute offers a full lesson plan around rotation using a special DIY apparatus to demonstrate that the rotational speed depends on the mass of the object that causes the centripetal acceleration (20).
> - **Gravitational lensing:** Probably the most famous analogy of gravitational lensing uses the base of a wine glass to distort an image (21,22). Moreover, Su (23) present a 3D-printable model to produce a plastic lens with a similar shape. Ros (21) also use a filled wine glass to demonstrate distortion effects.
> - **Curved spacetime:** To explain gravitational lensing, we need to introduce students to spacetime curvature. A popular analogy compares the distortion of spacetime to the distortion of a two-dimensional rubber sheet by massive objects (15,24).

Of course, all these activities can only serve as models and analogies of certain aspects of dark matter research. Physics educators should be aware of common conceptual difficulties and tailor the use of instructional models to the needs of their students. In line with (15), we understand analogies as one particular form of instructional models in science education that construct a similarity between two objects to aid interaction with and visualisation of science concepts. For example, the rubber sheet analogy compares the fabric of spacetime to a malleable rubber sheet. Kersting and Steier (15) recommend addressing strengths and weaknesses of this rubber sheet analogy explicitly when teaching general relativity. In a similar vein, Huwe and Field (22) introduce the wine glass model to demonstrate gravitational lensing while also encouraging learners to think about the role of the observer as well as limitations of the model. In contrast to this careful treatment of Huwe and Field, the wine glass model activities described by Ros (21) might even reinforce misconceptions about optical images.

Let us briefly look at two of the weaknesses of the wine glass model: First, the object that is observed through the base of the wine glass is often a printed image of a galaxy or a grid, not a light source. Therefore, a correct reconstruction of the image that avoids enforcing misconceptions about the process of seeing would require a light source and reflection of light at the printed image. Second, the wine glass model presents students with a three-dimensional lens. Yet, although geometric optics is often part of the physics curriculum, reconstruction of images and the effect of lenses is usually only discussed in two dimensions.

## 3. A fun approach to discuss dark matter in the classroom: jelly lenses

In the following, we present a fun and novel approach to teaching the basic principles of the gravitational effect of dark matter on light by using triangular jelly lenses. This jelly lens analogy used with a laser pointer reduces the lensing effect to two dimensions and, thus, eases the link to the existing physics curriculum. Indeed, the analogy allows students to apply their knowledge of optics, including the ray model of light and how to reconstruct images. We now turn to the presentation of the classroom activity. Doing so, we compare the jelly lens analogy to gravitational lenses, highlight the limitations of our analogy, and address well-known student conceptions in geometric optics that might interfere with the learning process.

### *3.1. Material*

The material listed in Box 3 is sufficient to make jelly lenses of different shapes for one or two groups of students. Box 4 suggests a unique gimmick for dark matter day.

> **Box 3: List of material**
>
> - A laser pointer that is safe to use.
>   **Careful**: Students should only use it under adult supervision and never direct it into someone's eye.
> - Food colouring in the colour of your laser pointer (e.g., red for red laser pointers)
> - 10 g gelatin
> - 500 ml water
> - 50 g sugar (or sweetener)
> - Optional: flavour
> - A small pot and a heater
> - Plastic or glass boxes, ideally in different shapes
> - A knife
> - Hot water
> - A surface you can clean easily such as a cutting board
> - A sheet of white paper

> **Box 4: A special fluorescent feature**
> To add a twist to your jelly lenses, get a UV lamp and use tonic water instead of water in the recipe. In this way, the jelly will get a unique taste, and under UV light, the jelly will fluoresce faintly in blue because of the quinine content of tonic water. We use this as a Halloween feature on dark matter day to point out the difference between the electromagnetically interacting jelly lens and dark matter lenses.

*3.2. Step-by-step instructions*

Using jelly to teach geometric optics and the effect of optical lenses is not a new idea. Indeed, Bunton (25) already describes an "edible optics activity". Here, we include our preferred jelly lens recipe and provide tips and tricks that we collected along the way. You can also watch a tutorial of the lens-making process and a short demonstration on YouTube[1].

- First, check the instructions on your gelatin but then use at least twice the amount of gelatin that is recommended: we use 10 g of gelatin together with half a litre of liquid. We tried producing a vegan jelly lens version using agar-agar instead of gelatin. However, this jelly does not become translucent enough for optics experiments.
- Next, add gelatin, water and sugar to a pot. Bunton (25) introduced the idea that the amount of sugar might change the refractive index of the jelly lens. However, there seems to be no easy linear relationship between the refraction index of a liquid and its density, which explains why we failed in producing different refractive indices.
- Add a few drops of food colouring to give the jelly a faint colour. Use the same colour as your laser point; this will increase the visibility of the laser beam. However, don't add too much colour; this could limit the penetration depth of your laser beam.
- If you want, you can add any kind of translucent flavours such as lemon juice or raspberry flavour. Yum!
- Gently heat the mixture and stir it until the gelatin and sugar are dissolved. There is no need to wait for the gelatin to soak up liquid as recommended in many recipes.
- Fill the jelly mix into plastic or glass containers and let the jelly solidify in a fridge for a few hours, ideally overnight. Make sure that the height of the jelly in the respective container is about 2 to 3 cm. Higher lenses will suffer from poor stability; lenses with a lower height might break easily.

---

[1] https://www.youtube.com/watch?v=MumHL3OC-kEfeature=youtu.be

- Once the jelly is solid enough, prepare a clean workspace, especially if you plan to eat the jelly after the experiment. Cooking spray or any type of oil can prevent the jelly from sticking to your surface, but it significantly affects its taste.
- To release the jelly from the different containers, put them into a bath of hot water for a few seconds. Also, put your knife into the hot water; this will make it easier to cut out jelly shapes.
- Cut out triangular jelly shapes using the knife. The shapes should represent isosceles triangles with an edge length of 5 to 10 cm.
- Fold a sheet of white paper twice lengthwise to serve as a screen.
- Place the triangular shape on a cutting board or similar, the two equal sides facing the screen, the other side facing the laser pointer.
- Finally, switch on your laser pointer and dim the lights to improve the visibility of the laser beam passing the jelly (Figure 3).

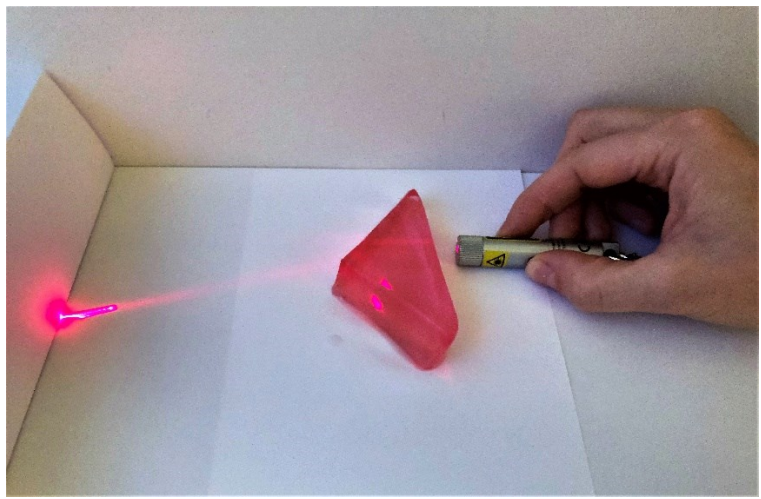

**Figure 3.** Refraction of red light emitted from a laser pointer by a triangular jelly lens.

## 3.3. The use and limitations of this analogy

*Elements of this analogy* In this analogy, the laser pointer represents light from a distant galaxy. The laser pointer can be directed at the jelly lens at different angles to represent multiple light rays coming from the light source. The folded sheet of white paper represents the observer, for example, a telescope on Earth. The triangular jelly shape represents a gravitational dark matter lens, and the jelly itself represents curved spacetime.

*Effect of jelly versus dark matter on light* If no jelly is placed between the laser pointer and the screen, light appears to travel in a straight line. However, if the laser light passes through the jelly, it changes direction. Similarly, if light passes through spacetime that is curved by dark matter, it looks like the light changes direction. In particular, light rays appear to be bent around dark matter clusters.

*Can we see light beams from the side?* As described in section 2.2, some students think that we can see light beams from the side. Many visualisations of gravitational lensing use a ray model of light suggesting that we can see the light ray from the side as they are bent around a massive object. This might reinforce this misconception. Therefore, we recommend addressing the visibility of laser light beams, which is easily demonstrated with a laser pointer traversing air. We can only make a laser beam visible if it passes through a translucent medium, such as jelly, that forces some laser light to scatter into our eyes. Similarly, an observer on Earth cannot see the trajectories of light rays and will not know that the light passed through curved spacetime.

*The role of the observer* An observer cannot see the original light source without the lens. Therefore, assuming that light travels in straight lines, an observer will infer a different location of the light source based on what they can see: they would reconstruct an image of the light source. Similarly, an observer on Earth will infer a different location of a distant galaxy in case of gravitational lensing effects.

It can be challenging to understand the role of the observer and their position in the creation of images both for optical lensing and gravitational lensing. Therefore, we recommend letting students simulate different configurations of the light source, the lens and the observer, for example, by using the Ray Optics Simulation[2]. In this simulation, students can both create lenses of different shapes with varying refractive indices and change the light source, including the density of light rays and the position of the observer (Figure 4). Working with this simulation, educators can explicitly address student conceptions related to the image reconstruction process, in particular, when using triangular lenses.

---

[2] https://ricktu288.github.io/ray-optics/simulator/

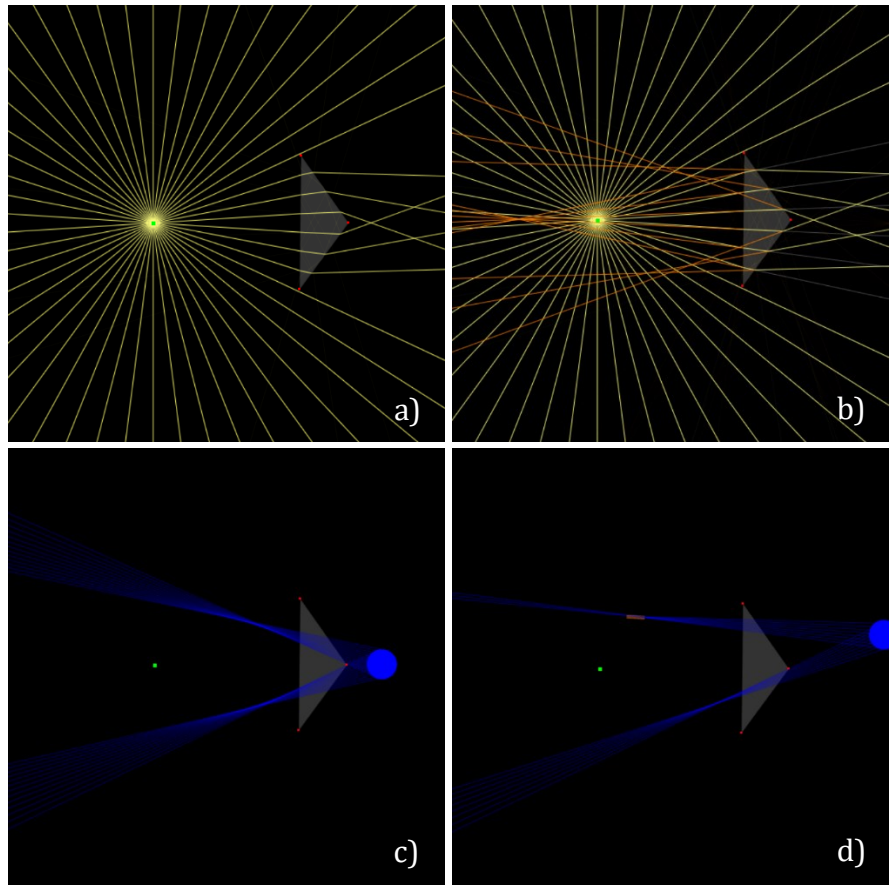

**Figure 4.** Screenshots of different configurations of the Ray Optics Simulation[2] with a triangular lens: a) light ray representation, b) extended light ray representation, c) example of an observer and respective images, d) second example of an observer and respective images.

*Why a triangular shape?* Converging lenses are symmetrical and have a focal point, which leads to one sharp of a light source. The further away from the optical axis light traverses the lens, the higher the refraction angle. Gravitational lenses, on the other hand, are usually not symmetrical and do not have a focal point. Instead, the further away from the centre of the dark matter mass light passes, the smaller the spacetime curvature and the smaller the bending effect on light. The index of refraction of space at a certain point is proportional to the gravitational potential there. Therefore, gravitational lenses have a focal line rather than a focal point. Consequently, gravitational lensing effects lead to displacement and distortions of the observed images, but only very rarely to magnifications. Indeed, in the early years of research on gravitational effects on light, authors used the term "gravitational mirages" instead of "gravitational lensing effect", in particular, if several images of a single source were produced (26).

In the jelly lens analogy, a triangular lens is used because it does not have a focal point either. Instead, the size of the angle of refraction only depends on the shape of the triangle. No matter how far away from the optical axis light traverses the jelly lens, the refraction angle will always be the same. The jelly lens is not an exact analogy to gravitational lensing; however, it does convey the main differences to a converging lens. Also, it is relatively easy to cut out triangular jelly shapes.

*Positions and alignment of the light source, lens, and observer* Depending on the distances between the light source, the lens, and the observer and their alignment, different images will be reconstructed, both for gravitational lenses and for jelly lenses. Here, simulations easily allow comparing different configurations. Einstein rings are one example of the effect of alignment. Suppose the light source, a point-mass gravitational lens, and an observer are perfectly aligned. In that case, the image of the light source appears as a circle of light, a so-called Einstein ring. However, if the gravitational lens is extended in size or the alignment is not perfect, the observers might see multiple, potentially distorted images.

## 4. Summary and conclusion

Dark matter is a fascinating topic not just for today's scientists but also for high school teachers and students. It offers the opportunity to engage and motivate students while also allowing discussions about crucial aspects of the nature of science. Even more, dark matter allows placing knowledge of geometric optics in an exciting context.

In this article, we explored the use of jelly lenses as a fun (and tasty!) new instructional analogy of gravitational lenses. This activity allows discussing the phenomenon of gravitational lensing in two dimensions, and thus, can form the basis of more complicated models such as the wine glass model. We have framed our instructional approach in line with the Model of Educational Reconstruction. Our presentation, therefore, provides a useful contextualisation of the subject matter for teachers who wish to learn more about the science content, student conceptions and different instructional resources of dark matter.

In the context of dark matter day, we would like to encourage our readers to explore the Halloween feature proposed in Box 4. Fluorescence is an elegant reminder of the difference between the electromagnetically interacting jelly lens and dark matter lenses. Bon appétit!